\documentclass[12pt]{article}
\usepackage{float}
\usepackage{amsmath}
\usepackage{amsfonts}
\usepackage{amssymb}
\usepackage{graphicx}
\usepackage{cite}
\usepackage{color}
\usepackage{dcolumn}
\usepackage{wrapfig}
\usepackage{longtable}
\usepackage{times}
\usepackage{caption}
\usepackage{subcaption}
\usepackage{hyperref}
\usepackage{bm}
\usepackage[margin=0.7in]{geometry}
\usepackage{newfloat}
\DeclareFloatingEnvironment[name={SM Figure}]{suppfigure}
\DeclareFloatingEnvironment[name={SM Table}]{supptable}
\begin{document}
\begin{flushleft}
{\LARGE
\textbf{Delay engineered solitary states in complex networks}
}
\\
\vspace{0.3cm}
\bf Leonhard Sch\"ulen$^1$, Saptarshi Ghosh$^2$, Ajay Deep Kachhvah$^2$, Anna Zakharova$^{1,*}$ and Sarika Jalan$^{2,3,\ast}$
\\
\vspace{0.2cm}
\it ${^1}$ Institut f{\"u}r Theoretische Physik, Technische Universit\"at Berlin, Hardenbergstra\ss{}e 36, 10623 Berlin, Germany
\\
\it ${^2}$ Complex Systems Lab, Discipline of Physics, Indian Institute of Technology Indore, Khandwa Road, Simrol, Indore 453552, India
\\
\it ${^3}$ Discipline of Biosciences and Biomedical Engineering, Indian Institute of Technology Indore,  Khandwa Road, Simrol, Indore 453552, India
\\
${^{*}}$ Email: anna.zakharova@tu-berlin.de \\
${^{\ast}}$ Email: sarikajalan9@gmail.com
\end{flushleft}

\begin{abstract}
We present a technique to engineer solitary states by means of delayed links in a network of neural oscillators and in coupled chaotic maps. Solitary states are intriguing partial synchronization patterns, where a synchronized cluster coexists with solitary nodes displaced from this cluster and distributed randomly over the network. We induce solitary states in the originally synchronized network of identical nodes by introducing delays in the links for a certain number of selected network elements. It is shown that the extent of displacement and the position of solitary elements can be completely controlled by the choice (values) and positions (locations) of the incorporated delays, reshaping the delay engineered solitary states in the network.
\end{abstract}

%% \linenumbers

%% main text
%--------------------------------------------------------------------------
\section{Introduction.} 
Synchronization is one of the well explored phenomena in the network science due to its widespread applications in various fields of science and technology \cite{PIK01,BOC18}. 
In particular, in many natural and man-made systems it is important to maintain the synchronized state of the network. The examples range from cardiac and neuronal tissue pacemakers to power grid networks. On the contrary, synchronization can be associated with an undesired pathological state, for instance, in neural networks in the case of epileptic seizure \cite{ROT14,AND16}. Synchronization is also involved in the generation of pathological movements, e.g., resting tremor in Parkinson’s disease \cite{TAS98}. In this context it is especially relevant to study and control the mechanisms of transition from synchronized to desynchronized regimes. 

Dynamical scenario of transition from completely synchronized to completely irregular behavior in complex networks can involve different types of partial synchronization patterns. One of the most prominent examples which has recently gained much attention is a pattern called chimera states \cite{KUR02a,ABR04}. Alternatively, such a transition can occur via another partial synchronization pattern, namely, \textit{solitary state} that in comparison to chimera states has been much less studied. Its mechanism is different from that of a chimera state. The term ``solitary'' comes from the Latin ``solitarius'' and can be understood as ``alone'', ``lonely'', or ``isolated''. In the case of chimera states the network spontaneously splits into coexisting domains of coherent (e.g., synchronized) and incoherent behavior, which are localized in space. On the contrary, for solitary states it is typical that individual ``solitary'' oscillators start leaving the synchronous cluster at random positions in space \cite{MAI14a,JAR15,MIK18}. Solitary states represent a soft transition to incoherence since the oscillators are leaving the coherent cluster gradually. Chimera states are characteristic of a sharp transition due to the fact that the occurring incoherent domain has initially a certain finite (typically large) size. Solitary states have various applications, for example, in power grid networks they may indicate a blackout \cite{HEL18}. They have been found in globally coupled ensembles of time-periodic van der Pol oscillators and chaotic R{\"o}ssler systems with attractive and repulsive interactions \cite{MAI14a}. Further, these states have been reported for locally, nonlocally and globally coupled networks of Kuramoto oscillators with inertia \cite{JAR15,JAR18}. Moreover, solitary states have been observed for chaotic maps \cite{RYB17,RYB18,SEM18a} and may play a role for neural networks \cite{MIK18}. Typically, solitary states occur due to the bistability and coexist with the synchronized solution of the system \cite{JAR18,SEM18a}. The investigation of partial synchronization patterns \cite{POE15,KRI17} is relevant for the understanding of the global dynamics in networks \cite{DO12}.

%look up SEM18a Intro for references on solitary states

Time delays can substantially influence emergent dynamics in coupled networks such as synchronization \cite{HAK06}, oscillation suppression \cite{ZAK13a,GJU14} and chimera states  \cite{GJU17,SAW17,SEM15b,ZAK17a}.  
They may be introduced artificially by external feedback loops and also arise naturally in many optical, electronic, neuronal or technological systems due to finite signal transmission and processing times, as well as memory and latency effects \cite{ATA10}. Examples are provided by delayed coupling or delayed feedback in coupled lasers, sensor networks, electronic circuits, power grids, communication and logistic networks. Both time-delayed coupling and feedback open up powerful methods of designing and controlling nonlinear dynamical systems by delay \cite{KAC19,POP11,YAN11,KAN15}. 

While previous works have been mainly devoted to the role of different delay types for the engineering chimera states \cite{GHO18a,GJU17} the impact of time-delayed couplings on solitary states is not yet well understood. Recently a scheme to design chimera states with the incoherence region depending on the distribution of delays in the sequential nodes has been reported \cite{GHO18a}. Here, we adopt the method to develop a novel scheme for forming a tailor-made solitary states by introducing delays in some edges to perturb certain nodes out of the coherent cluster.
We show that the fraction of nodes separating from the synchronized cluster can be controlled by implementing time delays. Focusing on a single layer network, we present a scheme to design solitary states using delays and apply it successfully to two different types of systems: time-discrete chaotic maps and FitzHugh-Nagumo systems.  This indicates a general, probably universal desynchronization mechanism in networks of very different nature that can be induced by adding time delays in the interactions.

\section{Model and Technique.}\label{sec2}

\begin{figure}
	\centerline{\includegraphics[width=\columnwidth]
		{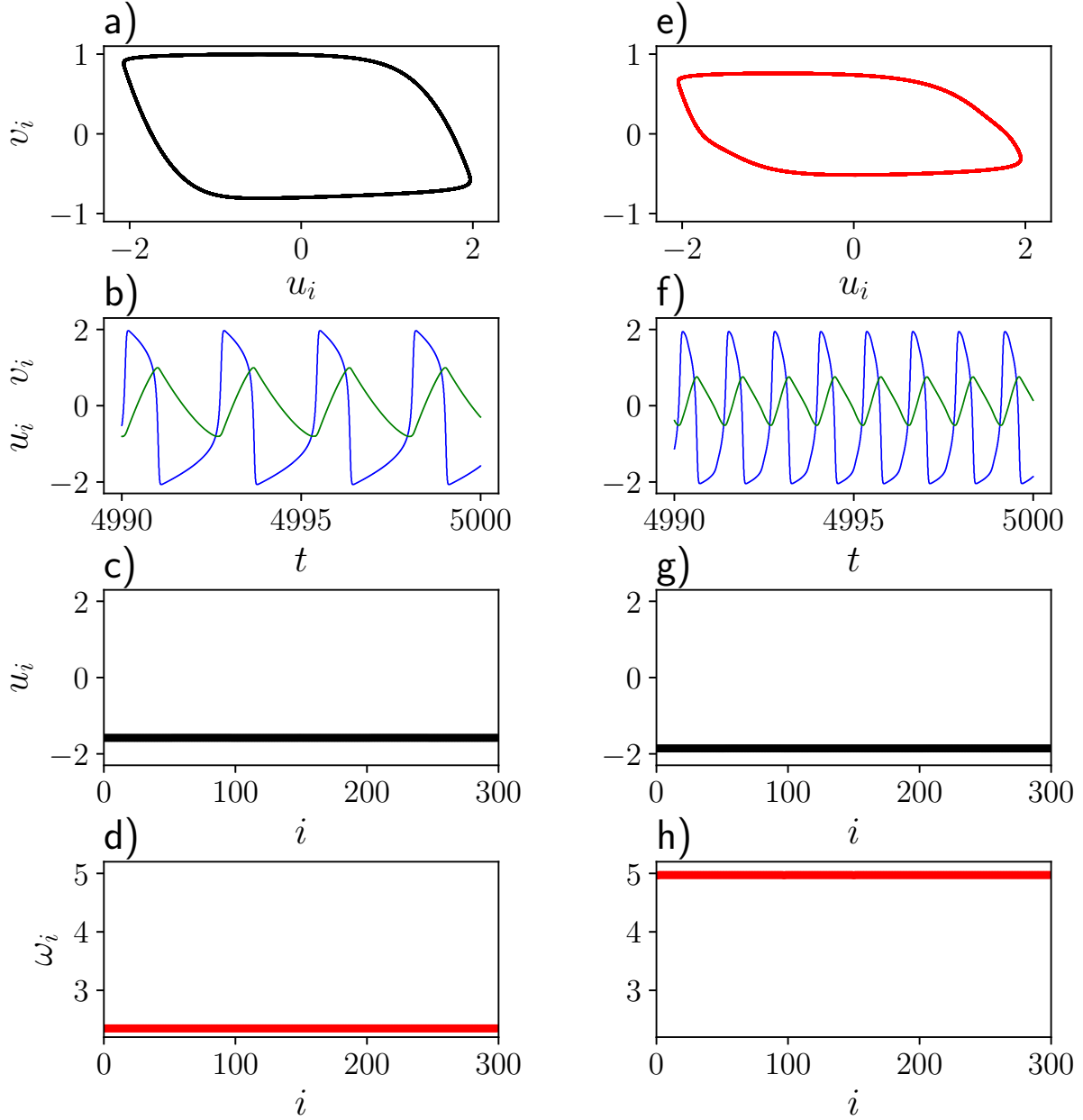}}
	\caption{(Color online) Comparison of a network Eq.~\ref{FHN} with no delay $\tau_{ij}=0$ (a-d) and all-delayed links $\tau_{ij}=1$ (e-h). Phase portraits (a,e), time series (b,f) (blue indicates activator $u_k$ and green shows inhibitor $v_k$), snapshots of the activator variables $u_k$ (c,g) and mean phase velocity profiles (d,h) are shown. Other parameters: $\varepsilon=0.05$, $a=0.5$, $N=300$, $\phi=\pi/2-0.35$, $r=0.35$, $\lambda=0.1$.}
	\label{fig8}
\end{figure}

In this section, we demonstrate the existence of solitary states as a consequence of the presence of delays in the network. The local dynamics of networked nodes is represented by FitzHugh-Nagumo (FHN) systems in oscillatory regime. This two-dimensional system is a paradigmatic model for neural excitability \cite{MAS17}.
Previously, in this model chimera states have been found for one-layer networks consisting of coupled oscillatory \cite{OME13,CHO18} and excitatory \cite{SEM16} FHN systems. 
Recently, this model has also been studied for two-layer networks: it has been shown that weak multiplexing plays a significant role for coherence resonance in ensembles of coupled excitable FHN units under the influence of noise \cite{SEM18}. The impact of time delays on chimera states has been investigated for one-layer \cite{SAW19}, two-layer \cite{NIK19}, and three-layer networks of FHN oscillators \cite{SAW19a}. Time-delayed feedback control of chimeras has been demonstrated for one-layer networks of FHN systems in excitable regime \cite{ZAK17a}. Solitary states in neural networks have been much less studied. Although the occurrence of solitary states for FHN model has been recently reported for two-layer networks of oscillatory FHN neurons \cite{MIK18}, the formation of these states in neural networks and, in particular, in the presence of time delays still remains to be understood. 

Here we consider a set of $N$ identical FHN oscillators arranged in a non-local ring fashion, each one of them connected to equal number of neighbors $R$ on either side of it. 
To incorporate delays in the system, a few nodes are selected at random and identical or non-identical delays $\tau_{ij}$ are then installed in $2R$ links of each selected node. Each such node is referred to as delayed node for the sake of convenience. Hence, matrix $\tau$ of delays comprises elements $\tau_{ij}$ such that $\tau_{ij}\neq0$ if time delay exists in a link between the nodes $i$ and $j$, and $\tau_{ij}=0$ otherwise. The entries $\tau_{ij}$ of the delay matrix depend on whether we consider homogeneous or heterogeneous delayed interactions for the delayed nodes. In instance of homogeneous delays, the entries $\tau_{ij}$ take the same value $\tau$ for all the delayed links of the network. In another instance of heterogeneous delays, the entries $\tau_{ij}$ are drawn from a random uniform distribution in a range $\tau_{ij}\in (0, \tau_{max}]$.
%Note that a delayed node means the node with all the edges originating from it are delayed. 
Hence, time-evolution of the dynamical state of the oscillators having either non-delayed or delayed interactions with $2R$ neighbors with homogeneous coupling strength $\sigma$ is given by
\begin{equation}\
\label{FHN}
\begin{array}{c}
\varepsilon\frac{du_i}{dt}=u_i-\frac{u^3_i}{3}-v_i 
+\frac{\lambda}{2P}\sum\limits_{j=i-P}^{i+P} b_{uu} (u_j(t - \tau_{ij}) -u_i(t)) \\
+ b_{uv} (v_j(t - \tau_{ij}) - v_i) \; ,\\
\frac{dv_i}{dt}=u_i + a + \frac{\lambda}{2P}\sum\limits_{j=i-P}^{i+P}
b_{vu} (u_j(t - \tau_{ij}) - u_i(t)) \\ 
+ b_{vv} (v_j(t - \tau_{ij}) - v_i(t)) \; ,
\end{array}
\end{equation}
where $u_i$ and $v_i$ are the activator and inhibitor variables, correspondingly, and $i=1,..,N$. A small parameter responsible for the time scale separation of fast activator and slow inhibitor is given by $\varepsilon > 0$. Here we fix $\varepsilon=0.05$. Parameter $a$ defines the excitability threshold. For an individual FHN element it determines whether the
system is in excitable $(|a| > 1)$, or oscillatory $(|a| < 1)$
regime. In this paper we consider the oscillatory regime and fix $a=0.5$. Our model includes not only direct, but also cross couplings between activator $u_i$ and inhibitor $v_i$ variables, which is modeled by a rotational coupling matrix:
\begin{equation}
B = \begin{pmatrix}
b_{uu} & b_{uv} \\
b_{vu} & b_{vv}
\end{pmatrix}
=
\begin{pmatrix}
cos \phi & sin \phi \\
-sin \phi & cos \phi
\end{pmatrix}.
\end{equation}

\begin{figure*}[t]
	\centerline{\includegraphics[width=0.8\textwidth]{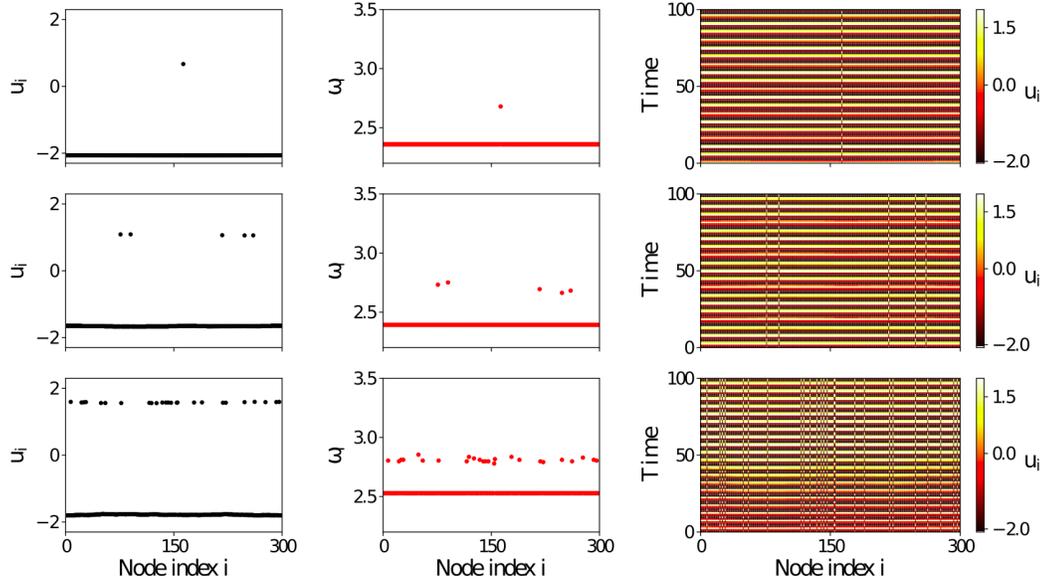}}
	\caption{(Color Online) Snapshots (left column), mean phase velocity profiles (middle column) and space-time plots (right column) for 1 (upper row), 5 (middle row) and 25 (bottom row) delayed nodes. The delay times are set to $\tau_{ij}=1$ for all selected nodes. Other parameters as in Fig. ~\ref{fig8}.} 
	\label{fig9}
\end{figure*}
Chimera states in system Eq.~\ref{FHN} for the undelayed coupling have been found for $\phi = \pi/2 - 0.1$ \cite{OME13}.
We fix $\phi=\pi/2 - 0.35$. For this value, complete synchronization is observed in the undelayed network $(\tau_{ij} = 0)$. In more detail, in the phase space we see the limit cycle (Fig.~\ref{fig8} a), the time evolution is periodic for all neurons (Fig.~\ref{fig8} b), the neurons are phase synchronized (Fig.~\ref{fig8} c) and all the units have the same mean phase velocity (Fig.~\ref{fig8} d). 

Qualitatively the same behavior is observed when all links in the network are homogeneously delayed with $\tau_{ij}=1$. The difference with respect to the undelayed case is that the amplitude of the inhibitor variable is smaller (Fig.~\ref{fig8} e, f) and the frequency is more than doubled (Fig.~\ref{fig8} f, h).
To demonstrate the solitary states, we consider a network of $N=300$ FHN oscillators interacting within the range of coupling radius $r=P/N=0.35$. The initial conditions for the oscillators are randomly distributed on a circle with $u^2 + v^2 = 4$. Next, we show the emergence of solitary states by adding delays in the links of individual oscillators.

\section{Delays as perturbation in un-delayed oscillators}
Here, we engineer solitary states for an undelayed network of identical FHN oscillators by introducing homogeneous delays for all ingoing and outgoing links of selected nodes. The solitary states are created by adding a homogeneous delay time $\tau_{ij}=1$ into all links of selected oscillators. We illustrate our results for the case of 1, 5 and 25 nodes having delayed links in Fig.~\ref{fig9}. In all cases, the solitary nodes have a slightly larger mean phase velocity (Fig.~\ref{fig9} middle column). Moreover, they influence the nodes from the coherent cluster whose frequency becomes larger, if compared to the undelayed case. 

\begin{figure}[t]
	\centerline{\includegraphics[width=0.8\columnwidth]
		{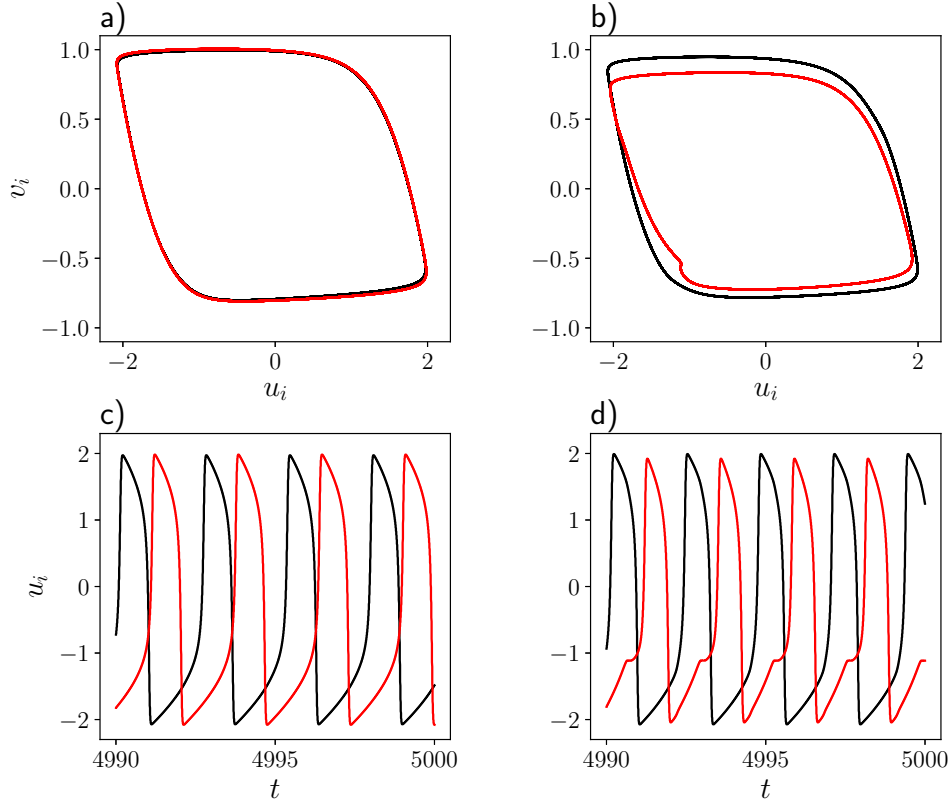}}
	\caption{(Color online) Phase portrait for one selected node from the synchronized cluster (black) and one selected solitary node (red) for 5 (a) and 75 (b) delayed nodes. Time series of the activator $u_i$ for the synchronized (black) and the solitary node (red) for 5 (c) and 75 (d) delayed nodes with $\tau_{ij} = 1$; other parameters as in Fig.~\ref{fig8}.}
	\label{fig10}
\end{figure}

Besides the changes in the mean phase velocity profile, we also observe a modification of the phase portrait. While for the case of 5 delayed nodes (Fig.~\ref{fig10} a) the limit cycle of the synchronized cluster is almost the same as that of the solitary nodes, the network with 75 delayed nodes is characterized by a smaller limit cycle for the solitary elements (Fig.~\ref{fig10} b). For both cases the system demonstrates periodic dynamics in time with solitary nodes phase-shifted from the elements belonging to the synchronized cluster (Fig. ~\ref{fig10} c, d). The smaller limit cycle can be explained as follows: the network with a few delayed nodes (i.e., nodes with all in-going and out-going links delayed) can be considered as an interpolation between two limit cases: a completely undelayed network characterized by a larger limit cycle (Fig.~\ref{fig8} a) and a network with all links delayed characterized by a smaller limit cycle (Fig. ~\ref{fig8} e). It is important to note, that previously a similar phase space structure (smaller limit cycle for solitary nodes) has been observed for a solitary state in the undelayed network of FHN oscillators \cite{MIK18}.

\begin{figure*}[t]
	\centerline{\includegraphics[width =\textwidth]
	{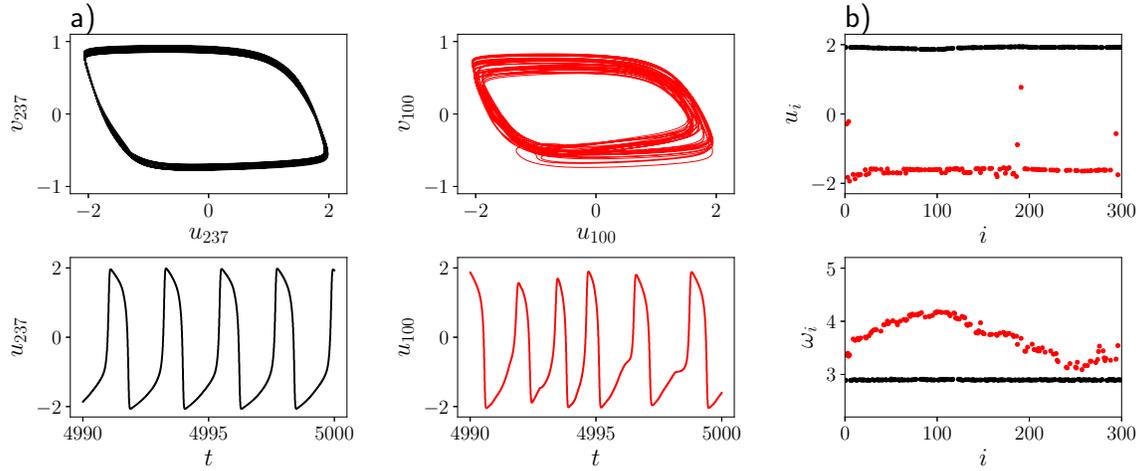}}
	\caption{(Color online) a) Comparison of the dynamics for one selected node in the synchronized cluster (black, left column) and that for one selected solitary node (red, right column). Top panels show the phase portraits, bottom panels illustrate the time series of the activator variable. b) Snapshot (top) and mean phase velocity profile (bottom) for the synchronized (black) and solitary (red) cluster; $\tau_{ij} = 1$; number of delayed nodes: $125$; other parameters as in Fig.~\ref{fig8}}.
\label{fig11}
\end{figure*}

If we increase the number of the solitary nodes further, the dynamics of the network becomes more complex. In Fig.~\ref{fig11} a) we separately show the behavior for one selected node from the synchronized cluster (Fig. ~\ref{fig11} a, left panel) and one selected solitary node (Fig. ~\ref{fig11} a, right panel). While the behavior of the synchronized cluster remains regular and time-periodic, the solitary node shows a much more complex dynamics. The limit cycle gets smeared out and the time series demonstrates irregular spiking of the FHN unit. The amplitude varies in time.
The more complex behavior in this case can also be seen from Fig.~\ref{fig11} b) where a snapshot of the activator $u_i$ and the mean phase velocity profile are shown. The black dots indicate the nodes from the synchronized cluster, whereas the red ones are the solitary nodes. Again, the mean phase velocity of all nodes is shifted by introducing delay. 

\begin{figure*}[t]
	\centerline{\includegraphics[width=\textwidth]{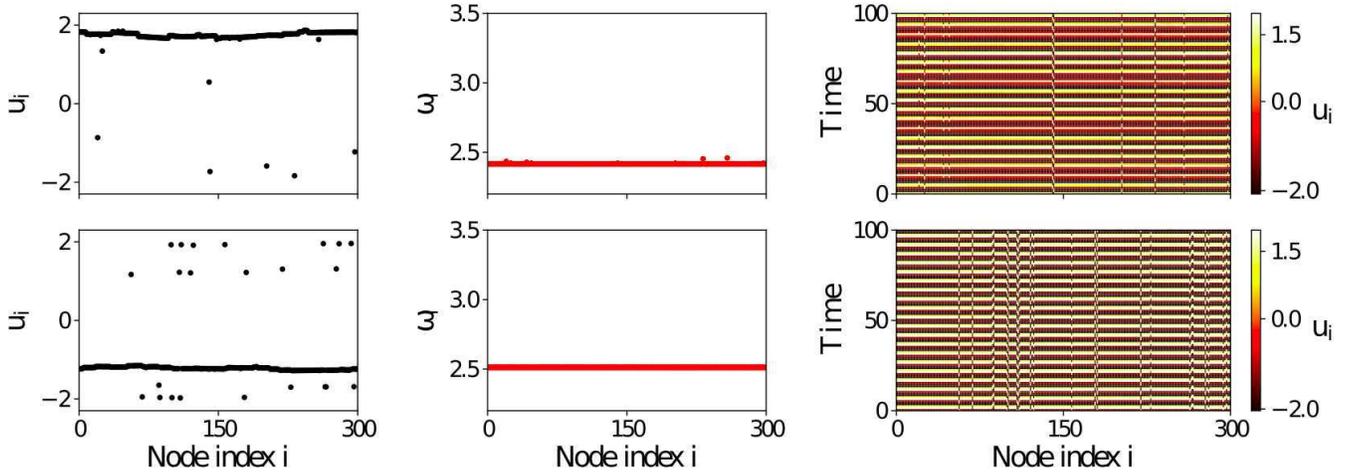}}
	\caption{(Color online) Snapshots of variable $u_i$ (left column), mean phase velocity profiles (middle column) and space-time plots (right column) for system Eq. ~\ref{FHN}, where randomly selected 10 (top panel) and 25 (bottom panel) nodes have delays randomly distributed over the interval $\tau_{ij} = [1, 10]$. Other parameters as in Fig. ~\ref{fig8}.}
	\label{fig12}
\end{figure*}

Next, we choose the delay times randomly in the interval $\tau_{ij}. \in [1, 10]$. Fig.~\ref{fig12} displays the snapshots, mean phase velocity profiles and the space-time evolution of the network with solitary states induced by introducing heterogeneous delay times. In the upper panels we delayed 10 nodes with random delay times within the interval, in the lower we did the same for 25 nodes. As opposed to the homogeneous case, the phase shift of the oscillators depends on the delay time. Strikingly the mean phase velocities show a rather flat profile.
This indicates frequency synchronization, different from the case of homogeneously delayed nodes. 

\section{Delays as perturbation in delayed oscillators}

\begin{figure}
	\centerline{\includegraphics[width=\textwidth]{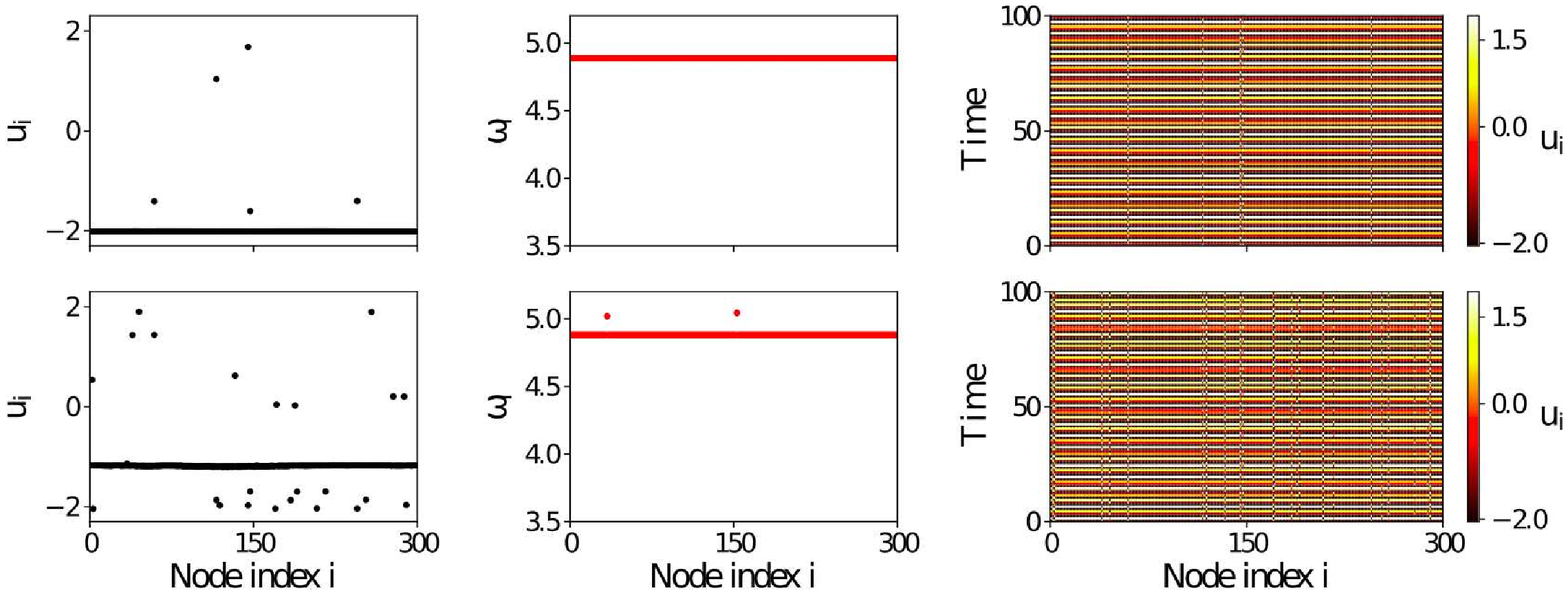}}
	\caption{(Color online) Snapshots of variable $u_i$ (left column), mean phase velocity profiles (middle column) and space-time plots (right column) for system Eq. ~\ref{FHN} where all nodes are delayed homogeneously with $\tau_{ij}=1$, except for randomly selected 5 (top panels) and 25 (bottom panels) nodes that have delays randomly distributed over the interval $\tau_{ij} = [2,10]$. Other parameters as in Fig. ~\ref{fig8}.}
\label{fig13}
\end{figure}

Next, we apply the reverse scheme: starting from a network of nodes with homogeneous delay times $\tau_{ij}=1$, we change the delay values for a small number of nodes and, thus, make the network heterogeneous. This is done by delaying a certain number of randomly selected nodes with delay times randomly chosen in the interval $\tau_{ij} \in [2, 10]$. The delay times influence the phase shift in the network (Fig. ~\ref{fig13}). As it is the case for the partially delayed network with heterogeneously delayed nodes, the mean phase velocity profile remains rather flat, with only small deviations. The upper panels in Fig. ~\ref{fig13} show the results for heterogeneously delaying 5 nodes and in the lower panels 25 nodes are heterogeneously delayed. Qualitatively the behavior is the same as in the partially delayed network with heterogeneous delays. We, therefore, conclude that the mechanism works in both directions.

\begin{figure}
	\centerline{\includegraphics[width=\columnwidth]{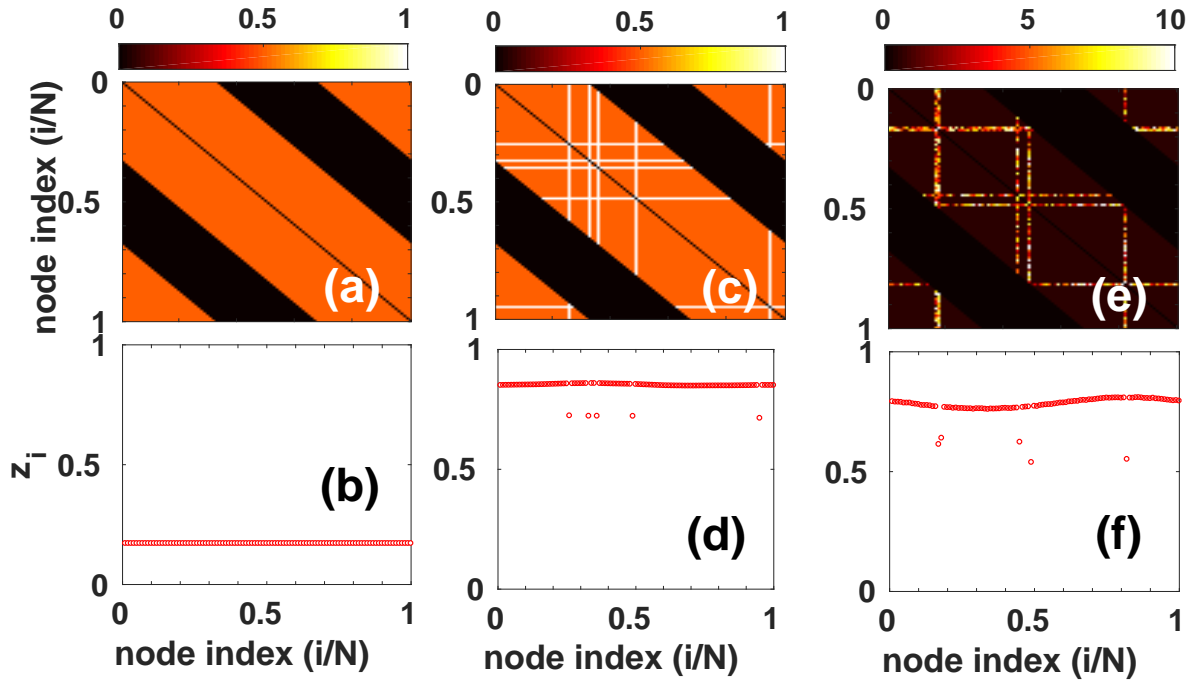}}
	\caption{(Color Online) Diagram depicts the composite-adjacency matrix of the delayed-undelayed links (upper panels) and corresponding spatial profiles (lower panels) for a few randomly selected undelayed $\tau_{ij}=0$ (left), homogeneously delayed $\tau_{ij}=1$ (middle), and heterogeneously delayed $\tau_{ij}\in(0,10]$ (right) nodes.}
	\label{fig1}
\end{figure}
\begin{figure}
  \centerline{\includegraphics[width=\columnwidth]{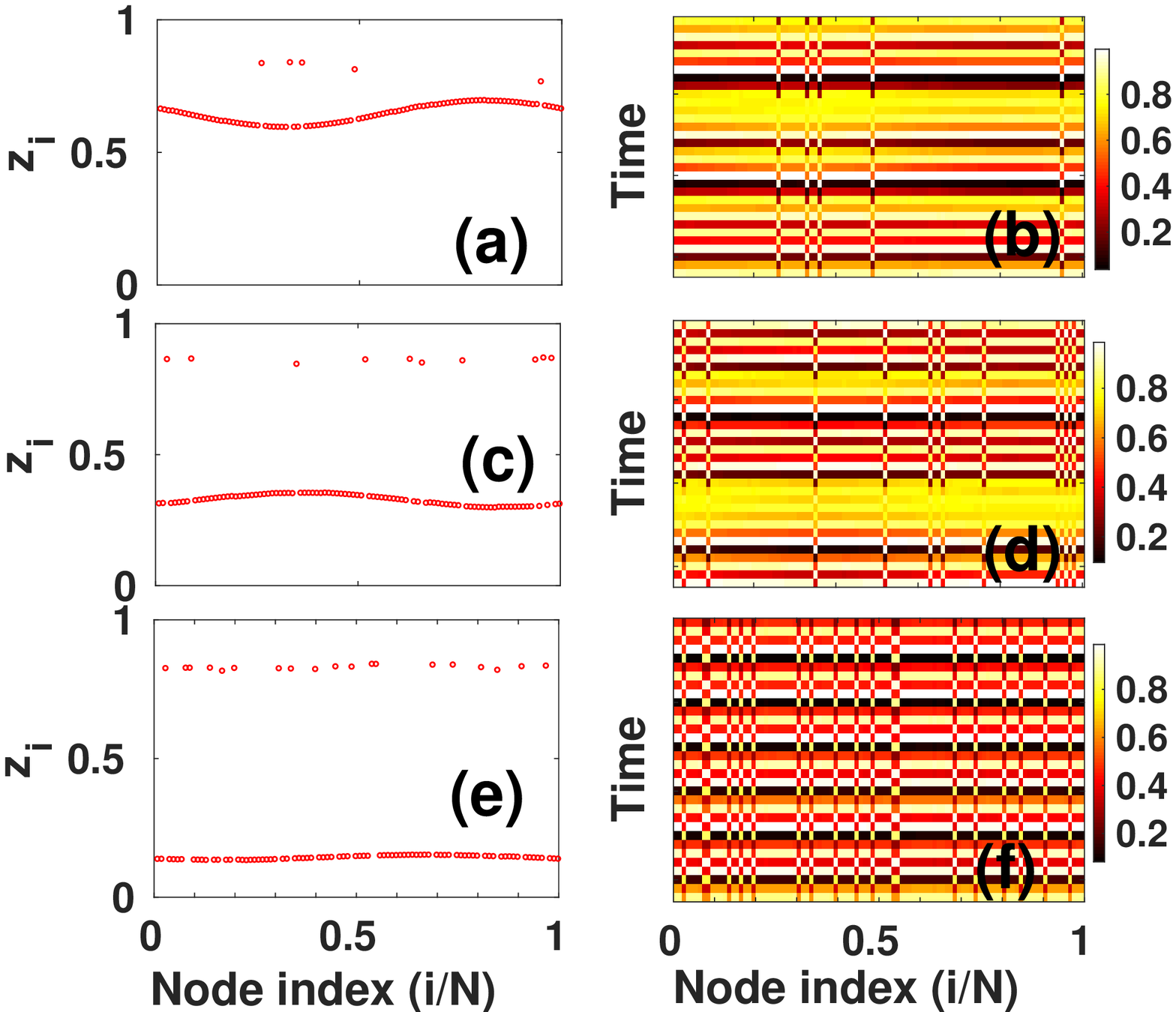}}
	\caption{(Color Online) Snapshots (left column) and spatial patterns (right column) of the solitary states generated from (a-b) $5$, (c-d) $10$ and (e-f) $20$ randomly chosen homogeneously ($\tau_{ij}=1$) delayed-nodes or set of delayed-edges.}
	\label{fig2}
\end{figure}

\section{Robustness of technique against maps}\label{sec5} 
Chaotic maps have been shown to demonstrate chimera states \cite{GHO16a,GHO16,GHO18} and solitary states \cite{RYB17,RYB18,SEM18a}. However, the impact of time delay on the formation of solitary states has not previously been investigated for coupled maps. Therefore, we approach this problem and demonstrate that the same scheme can be successfully employed to a network of a different nature, namely, chaotic time-discrete maps. To accomplish this, we again consider a ring network of $N$ identical nodes with $K$ neighbors either side of them. Now, the local dynamics of the nodes is characterized by time-discrete map $z(t+1)=f(z(t))$, where $f(z)=\mu z (1-z)$; $z\in[0,1]$ is a logistic map considered in its chaotic regime ($\mu=4.0$). Here the delayed nodes are chosen in the same fashion as before. Therefore, time-evolution of the dynamical state of each delayed or un-delayed node interacting with its neighbors with coupling strength $\varepsilon\in[0,1]$ is governed by
\begin{equation}
z_i(t+1)=f(z_i(t))+\frac{\varepsilon}{k_i} \sum_{j=1}^{N} A_{ij}[ f(z_j(t - \tau_{ij}))-f(z_i(t)) ].
\label{eq.devol}
\end{equation}
where $A_{ij}$ denotes the element of the adjacency matrix $A$ encoding regular topology, which takes vale 1 if $i_{th}$ and $j_{th}$ nodes are connected, and $0$ otherwise.  Therefore, $k_{i}$ = $\sum_{j=1}^{N}A_{ij}$ denotes the degree of the $i_{th}$ node. We define the common range of interaction of each node in the regular network as coupling radius $r=K/N=0.32$. The initial states of the nodes are selected uniform randomly in the range $z(t=0)\in[0,1]$. The coupling strength is kept fixed  $\varepsilon=0.9$ to make the system initially demonstrate synchronized state.

\subsection{Delays as perturbation in un-delayed maps}

To investigate the solitary states in the networked system, we perturb the synchronous system with the aid of either homogeneous or heterogeneous delays. 
In Fig.~\ref{fig1}, the upper panels portray the snapshots of the composite adjacency matrix $D$ of the delayed-undelayed links and the lower panels display corresponding spatial profiles for the un-delayed, homogeneously and heterogeneously delayed systems. Panels (a) and (b) display regular pattern of the undelayed links and the related completely coherent spatial states, respectively, for the undelayed network. Panel (c)  displays regular undelayed links with a few identical delayed ($\tau_{ij}=1$) links, and associated solitary states with almost equally displaced isolated nodes in panel (d) for the homogeneously delayed network. For the heterogeneously delayed network, in a similar fashion, panel (e) exhibits regular undelayed links with a few non-identical delayed ($0<\tau_{ij}\le10$) links, and corresponding solitary states with unequally displaced nodes in panel (f).

Next, we investigate the impact of number of delayed links $N_{\tau}$ on the emergent solitary states. Fig.~\ref{fig2} exhibits the spatial profiles and the corresponding snapshots of the solitary states as a corollary of different number of homogeneous delays installed in the system. It is quite obvious that the number of installed delays interpolates the equal number of solitary sites from the synchronous clusters. Thus, the number of delayed links $N_{\tau}$ provides us with a qualitative control over the emerging solitary states.

\subsection{Delays as perturbation in delayed maps}
\begin{figure}[t]
	\centerline{\includegraphics[width=0.8\columnwidth]{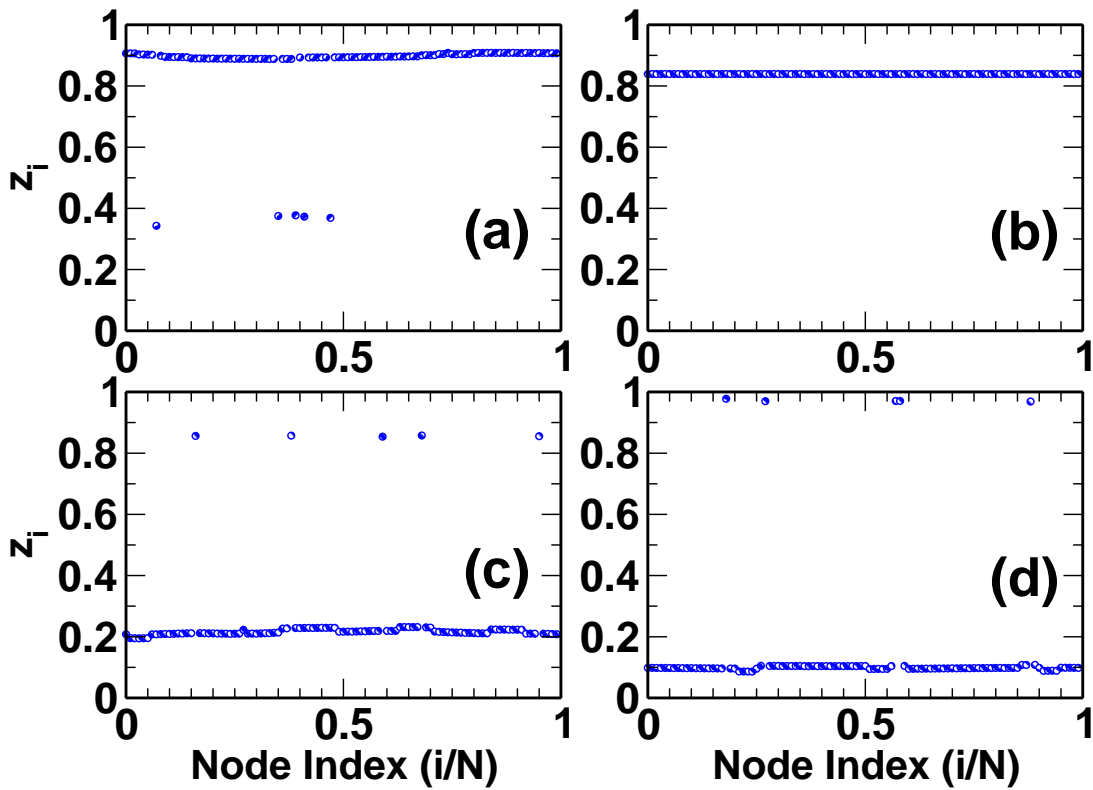}}
	\caption{(Color Online) Spatial profiles of the solitary states when initial synchronized cluster is generated from all homogeneously $\tau=1$ delayed nodes (b). The solitary states are interpolated by replacing few nodes with (a) $\tau_{ij}=0$, (c) $\tau_{ij}=2$ and (d) $\tau_{ij}=3$ nodes.} 
	\label{fig3}
\end{figure}

Now we employ a reverse scheme to achieve the initial synchronous state of the system by taking into account all the homogeneously delayed nodes, the synchronized state is then perturbed by replacing delay values in a few randomly selected delayed nodes with (i) a set of null ($\tau_{ij}=0$) or identical delays other than the existing ones, or (ii) a set of heterogeneous delays. The reverse scheme for the emergence of the solitary states is employed in Fig.~\ref{fig3} using (i) different identical delays. Panel \ref{fig3}(b) exhibits a completely synchronized state interpolated by homogeneously delayed nodes ($\tau_{ij}=1$). 
In the panels \ref{fig3}(a), \ref{fig3}(c) and \ref{fig3}(d), the solitary states appear as a corollary of replacing few existing delayed nodes with $\tau_{ij}=0$, $\tau_{ij}=2$ and $\tau_{ij}=3$ ones, respectively. The emergent solitary states are almost equally dislodged from the synchronized cluster.

\begin{figure}[t]
	\centerline{\includegraphics[width=0.8\columnwidth]{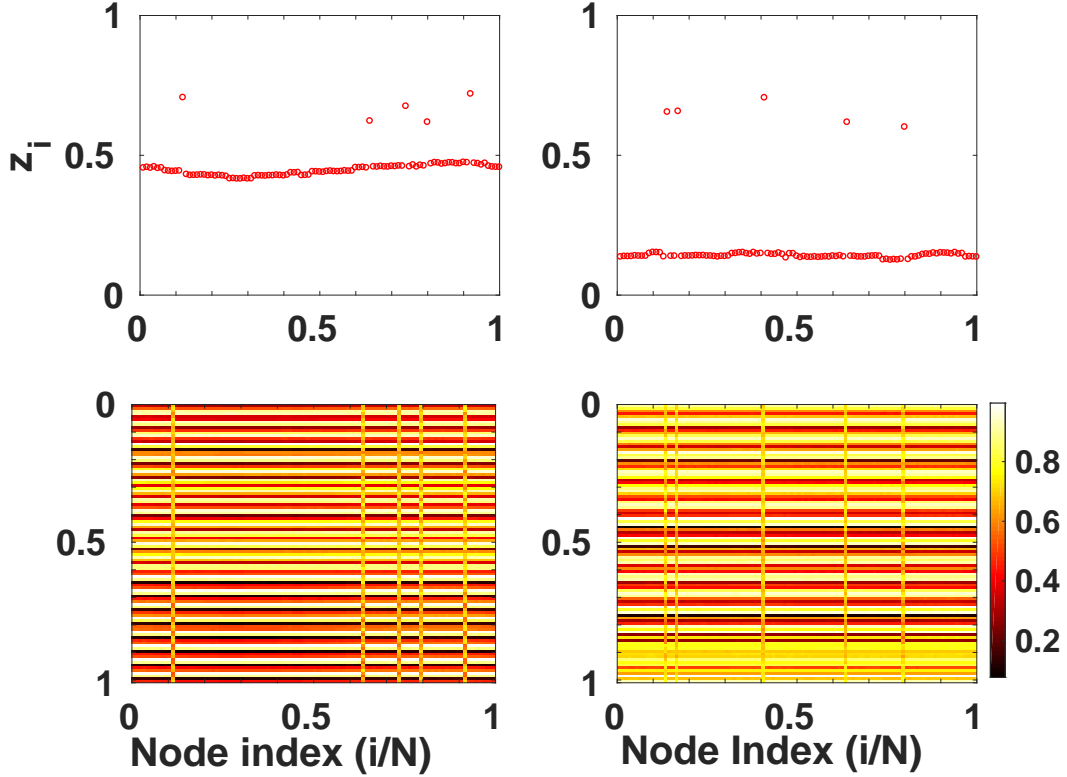}}
	\caption{(Color Online) Snapshots (top panels) and spatial profiles (bottom panels) of the solitary states when initial synchronized cluster is generated from all homogeneously delayed nodes (a) $\tau=1$ (left panels) and (b) $\tau=2$ (right panels). The solitary states are interpolated by replacing a few existing homogeneous delayed nodes with heterogeneous delayed-nodes in the range $\tau_{ij}\in(0,10]$.}
	\label{fig4}
\end{figure}
Now, we employ the reverse scheme for the emergence of the solitary states using (ii) heterogeneous delays in Fig.~\ref{fig4}. In Fig.~\ref{fig4}, the initial synchronous states are generated using two different sets of identical delays $\tau_{ij}=1$ (left panels) and $\tau_{ij}=2$ (right panels). Now, the solitary states are interpolated in the global synchronous state by replacing a few randomly selected existing delayed nodes with heterogeneously delayed ones such that $\tau_{ij}\in(0,10]$. It is quite apparent from the spatial profiles and snapshots that the emergent solitary states are unequally dislodged from the synchronous cluster.
Hence, even with the reverse scheme, one can construct both the equally and unequally dislodged solitary states by making appropriate choices of inducing delays.

\section{Conclusions.} 
In this work, we present a technique to engineer solitary states in a synchronous system by incorporating delayed nodes, i.e., installing a common delay in all the links of a node. We show that provided this technique one can construct solitary states in systems of oscillators. This method allows us to induce solitary states with identical and non-identical spatial patterns by installing homogeneous or heterogeneous delayed-nodes, respectively. Also, the extent of phase-displacement of the identical solitary nodes from the synchronous cluster can be controlled by making a proper choice of time delay(s). The employed technique also enables us to dislodge the desired number of solitary states at desired locations in the synchronized system. Thus, provided this method one can engineer the solitary states in a coherent system according to one's preferences. We demonstrate the robustness of the technique by constructing solitary states in maps with aid of the delayed nodes. Hence, the technique presented here in engineering solitary states is applicable to systems of different nature.\\

\section*{Acknowledgments}
SJ acknowledges DST project grant (EMR/2016/001921) for financial support. SG acknowledges DST Government of India for the INSPIRE fellowship (IF150149) and members of Complex Systems Lab for useful discussions. AZ and SJ acknowledge support from DST-DAAD PPP project (INT/FRG/DAAD/P-06/2018). LS and AZ acknowledge support by the Deutsche Forschungsgemeinschaft (DFG, German Research Foundation) - Projektnummer - 163436311 - SFB 910. 

%\bibliographystyle{apsrev}

%\bibliographystyle{unsrt}
%\bibliography{ref}

\end{document}